\documentclass[prl,twocolumn,superscriptaddress,longbibliography,preprintnumbers,nofootinbib]{revtex4-1}

\usepackage{url,xspace,dsfont,amssymb,amsmath,graphicx,multirow,hhline}
\usepackage[caption=false]{subfig}
\usepackage[colorlinks=true,citecolor=blue]{hyperref}
\usepackage{colortbl}

\usepackage{color}
\definecolor{darkred}{rgb}{1.0,0.1,0.1}
\definecolor{darkgreen}{rgb}{0.1,0.7,0.1}
\definecolor{darkblue}{rgb}{0.1,0.1,1.0}
\definecolor{darkpurple}{rgb}{0.4,0.1,0.8}
\definecolor{darkorange}{rgb}{1.0,0.5,0.0}

\DeclareRobustCommand{\Tab}[1]{Table~\ref{tab:#1}}

\DeclareRobustCommand{\Fig}[1]{Fig.~\ref{fig:#1}}

\DeclareRobustCommand{\Eq}[1]{Eq.~(\ref{eq:#1})}

\DeclareRobustCommand{\Ref}[1]{Ref.~\cite{#1}}
\DeclareRobustCommand{\Refs}[1]{Refs.~\cite{#1}}

\newcommand{\Pythia}{{\sc Pythia}\xspace}
\newcommand{\Herwig}{{\sc Herwig}\xspace}
\newcommand{\FastJet}{{\sc FastJet}\xspace}
\newcommand{\Delphes}{{\sc Delphes}\xspace}

\newcommand{\OmniFold}{{\sc OmniFold}\xspace}

\newcommand{\MultiFold}{{\sc MultiFold}\xspace}
\newcommand{\UniFold}{{\sc UniFold}\xspace}
\newcommand{\EnergyFlow}{{\sc EnergyFlow}\xspace}

\begin{document}

\title{OmniFold: A Method to Simultaneously Unfold All Observables}

\preprint{MIT-CTP 5155}

\author{Anders Andreassen}
\email{ajandreassen@google.com}
\affiliation{Department of Physics, University of California, Berkeley, CA 94720, USA}
\affiliation{Physics Division, Lawrence Berkeley National Laboratory, Berkeley, CA 94720, USA}
\affiliation{Google, Mountain View, CA 94043, USA}

\author{Patrick T. Komiske}
\email{pkomiske@mit.edu}
\affiliation{Center for Theoretical Physics, Massachusetts Institute of Technology, Cambridge, MA 02139, USA}

\author{Eric M. Metodiev}
\email{metodiev@mit.edu}
\affiliation{Center for Theoretical Physics, Massachusetts Institute of Technology, Cambridge, MA 02139, USA}

\author{Benjamin Nachman}
\email{bpnachman@lbl.gov}
\affiliation{Physics Division, Lawrence Berkeley National Laboratory, Berkeley, CA 94720, USA}

\author{Jesse Thaler}
\email{jthaler@mit.edu}
\affiliation{Center for Theoretical Physics, Massachusetts Institute of Technology, Cambridge, MA 02139, USA}

\begin{abstract}
Collider data must be corrected for detector effects (``unfolded'') to be compared with many theoretical calculations and measurements from other experiments.
Unfolding is traditionally done for individual, binned observables without including all information relevant for characterizing the detector response.
We introduce \OmniFold, an unfolding method that iteratively reweights a simulated dataset, using machine learning to capitalize on all available information.
Our approach is unbinned, works for arbitrarily high-dimensional data, and naturally incorporates information from the full phase space.
We illustrate this technique on a realistic jet substructure example from the Large Hadron Collider and compare it to standard binned unfolding methods.
This new paradigm enables the simultaneous measurement of all observables, including those not yet invented at the time of the analysis.
\hspace{-.5in}
\end{abstract}

\maketitle

%
Measuring properties of particle collisions is a central goal of particle physics experiments, such as those at the Large Hadron Collider (LHC).
After correcting for detector effects, distributions of collider observables at ``truth level'' can be compared with semi-inclusive theoretical predictions as well as with measurements from other experiments.
These comparisons are widely used to enhance our understanding of the Standard Model, tune parameters of Monte Carlo event generators, and enable precision searches for new physics.
``Unfolding'' is the process of obtaining these truth distributions ({particle-level}) from measured information recorded by a detector ({detector-level}).
The unfolding process ensures that measurements are independent of the specific experimental context, allowing for comparisons across different experiments and usage with the latest theoretical tools,\footnote{For fully exclusive theoretical predictions, one could alternatively forward fold to compare to experimental data.} even long after the original analysis is completed.
Many unfolding methods have been proposed and are currently used by experiments.
See \Refs{Cowan:2002in,Blobel:2203257,doi:10.1002/9783527653416.ch6,Balasubramanian:2019itp} for reviews and \Refs{DAgostini:1994fjx,Hocker:1995kb,Schmitt:2012kp} for the most widely-used unfolding algorithms.

%
Current unfolding methods face three key challenges.
First, all of the widely-used methods require the measured observables to be binned into histograms.
This binning must be determined ahead of time and is often chosen manually.
Second, because the measurements are binned, one can only unfold a small number of observables simultaneously.
Multi-differential cross section measurements beyond two or three dimensions are simply not feasible.
Finally, unfolding corrections for detector effects often do not take into account all possible auxiliary features that control the detector response.
Even though the inputs to the unfolding can be calibrated, if the detector response depends on features that are not used directly in the unfolding, then the results will be suboptimal and potentially biased.

%
This letter introduces \OmniFold, a new approach that solves all three of these unfolding challenges.
Detector-level quantities are iteratively unfolded, using machine learning to handle phase space of any dimensionality without requiring binning.
Utilizing the full phase space information mitigates the problem of auxiliary features controlling the detector response.
There have been previous proposals to use machine learning methods for unfolding~\cite{Gagunashvili:2010zw,Glazov:2017vni,Datta:2018mwd} as well as proposals to perform unfolding without binning~\cite{Aslan:2003vu,Lindemann:1995ut,Glazov:2017vni,Datta:2018mwd}.
These proposals, however, are untenable in high dimensions and do not reduce to standard methods in the binned case.
\OmniFold naturally processes high-dimensional features, in the spirit of previous machine-learning-based reweighting strategies~\cite{Martschei_2012,Rogozhnikov:2016bdp,Aaij:2017awb,Aaboud:2018htj,Andreassen:2018apy,Andreassen:2019nnm}, and it reduces to well-established methods~\cite{DAgostini:1994fjx} in the binned case.
We also introduce simpler versions of the procedure, using single or multiple observables, named \UniFold and \MultiFold, respectively.\footnote{The name \OmniFold\ is taken from Emily Dickinson's poem \textit{The Mountain Sat Upon the Plain}~\cite{dickinson}.}


All unfolding methods require a trustable detector simulation to estimate the detector response.
In the binned formulation, the folding equation can be written as ${\bf m} = {\bf R}\,{\bf t}$, where ${\bf m}$ and ${\bf t}$ are vectors of the measured detector-level and true particle-level histograms, respectively.
$\bf R$ is the ``response matrix'':
\begin{align}
    R_{ij} = \Pr(\text{measure $i$} \,|\, \text{truth is $j$}).
\end{align}
In general, ${\bf R}$ is not invertible, so the unfolding problem has no unique solution, and methods attempt to achieve a useful solution in various ways.
One of the most widely-used methods is Iterative Bayesian Unfolding (IBU)~\cite{DAgostini:1994fjx}, also known as Richardson-Lucy deconvolution~\cite{1974AJ.....79..745L,Richardson:72}.
Given a measured spectrum $m_i = \Pr(\text{measure $i$})$ and a prior spectrum $t_j^{(0)} = \Pr_0(\text{truth is $j$})$, IBU proceeds iteratively according to the equation:
\begin{align}
  \nonumber
  t_j^{(n)} & = \sum_i \text{Pr}_{n-1}(\text{truth is $j$} \,|\, \text{measure $i$})
 \Pr(\text{measure $i$}) \\
  &= \sum_i \frac{R_{ij} t_j^{(n-1)}}{\sum_k R_{ik} t_k^{(n-1)}} \times m_i,
  \label{eq:unfolding}
\end{align}
where $n$ is the iteration number.
%

%
\OmniFold uses machine learning to generalize \Eq{unfolding} to the unbinned, full phase space.
A key concept for this approach is the likelihood ratio:
\begin{align}
\label{eq:LLR}
L[(w,X),(w',X')](x) = \frac{p_{(w,X)}(x)}{p_{(w',X')}(x)},
\end{align}
where $p_{(w,X)}$ is the probability density of $x$ estimated from empirical weights $w$ and samples $X$.
The function $L[(w,X),(w',X')](x)$ can be approximated using a classifier trained to distinguish $(w,X)$ from $(w',X')$.
This property has been successfully exploited using neural networks for full phase-space Monte Carlo reweighting and parameter estimation~\cite{Andreassen:2019nnm,Cranmer:2015bka,Brehmer:2018kdj,Brehmer:2018eca,Brehmer:2018hga,Bothmann2019}.
Here, we use neural network classifiers to iteratively reweight the particle- and detector-level Monte Carlo weights, resulting in an unfolding procedure.

\begin{figure}[t]
    \centering
    \includegraphics[width=\columnwidth]{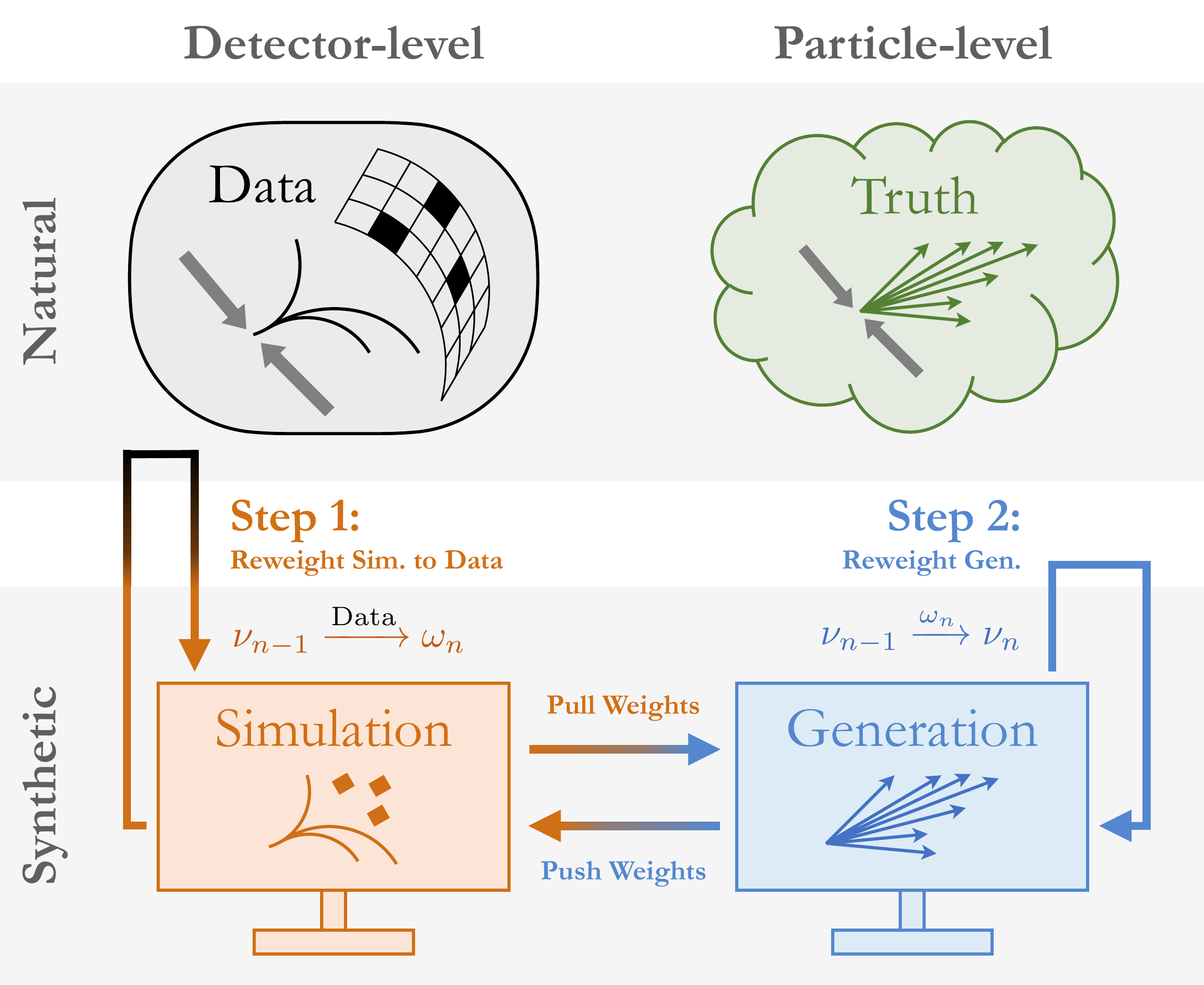}
    \caption{An illustration of \OmniFold, applied to a set of synthetic and natural data.
    As a first step, starting from prior weights $\nu_0$, the detector-level synthetic data (``simulation'') is reweighted to match the detector-level natural data (simply ``data'').  
    These weights $\omega_{1}$ are pulled back to induce weights on the particle-level synthetic data (``generation'').
    As a second step, the initial generation is reweighted to match the new weighted generation.
    The resulting weights $\nu_{1}$ are pushed forward to induce a new simulation, and the process is iterated.}
    \label{fig:method}
\end{figure}

%
The \OmniFold technique is illustrated in \Fig{method}.
Intuitively, synthetic detector-level events (``simulation'') are reweighted to match experimental data (``data''), and then the reweighted synthetic events, now evaluated at particle-level (``generation''), are further reweighted to estimate the true particle-level information (``truth'').
The starting point is a synthetic Monte Carlo dataset composed of pairs $(t,m)$, where each particle-level event $t$ is pushed through the detector simulation to obtain a detector-level event $m$.
Particle-level events have initial weights $\nu_0(t)$, and when $t$ is pushed to $m$, these become detector-level weights $\nu^{\rm push}_0({m})=\nu_0({t})$.
\OmniFold iterates the following steps:
\begin{enumerate}
    \item $\omega_{n}( m) = \nu^{\rm push}_{n-1}( m)\,L[(1,\text{Data}),(\nu^{\rm push}_{n-1},\text{Sim.})]( m)$,\label{step1}
    \item $\nu_{n}( t) = \nu_{n-1}( t)\,L[(\omega^{\rm pull}_{n},\text{Gen.}),(\nu_{n-1},\text{Gen.})]( t).$\label{step2}
\end{enumerate}
The first step yields new detector-level weights $\omega_{n}({m})$, which are pulled back to particle-level weights $\omega^{\rm pull}_{n}({t})=\omega_{n}({m})$ using the same synthetic pairs $(t,m)$.
Note that $\nu^{\rm push}$ and $\omega^{\rm pull}$ are not, strictly speaking, functions because of the multi-valued nature of the detector simulation.
The second step ensures that $\nu_{n}$ is a valid weighting function of the particle-level quantities.

Assuming $\nu_0(t)=1$, in the first iteration Step~\ref{step1} learns $\omega_1(m)=p_{\text{Data}}(m)/p_{\text{Sim.}}(m)$, which is pulled back to the particle-level weights $\omega^{\rm pull}_1(t)$.
Step~\ref{step2} simply converts the per-instance weights $\omega^{\rm pull}_1(t)$ to a valid particle-level weighting function $\nu_{1}(t)$.
After one iteration, the new induced truth is:
\begin{align}
\nu_1(t)\,p_{\text{Gen.}}(t) &= \int dm'\, p_{\text{Gen.}|\text{Sim.}}(t|m')\,p_{\text{Data}}(m').
\end{align}
This is a continuous version of IBU from \Eq{unfolding}, where the sum has been promoted to a full phase-space integral.
In fact, \OmniFold (and IBU) are iterative strategies that converge to the maximum likelihood estimate of the true particle-level distribution~\cite{shepp1982maximum,10.2307/2984875,wu1983,doi:10.1080/01621459.1985.10477119,Kuusela2012StatisticalII}, which we discuss in detail in the Appendix.
After $n$ iterations, the unfolded distribution is:
\begin{equation}
\label{eq:final_answer}
    p^{(n)}_\text{unfolded}(t)=\nu_n(t) \, p_{\text{Gen.}}(t).
\end{equation}
The unfolded result can be presented either as a set of generated events $\{{t}\}$ with weights $\{\nu_n({t})\}$ (and uncertainties) or, more compactly, as the learned weighting function $\nu_n$ and instructions for sampling from $p_\text{Gen.}$.

\begin{figure*}[t]
    \centering
    \includegraphics[height=0.305\textwidth]{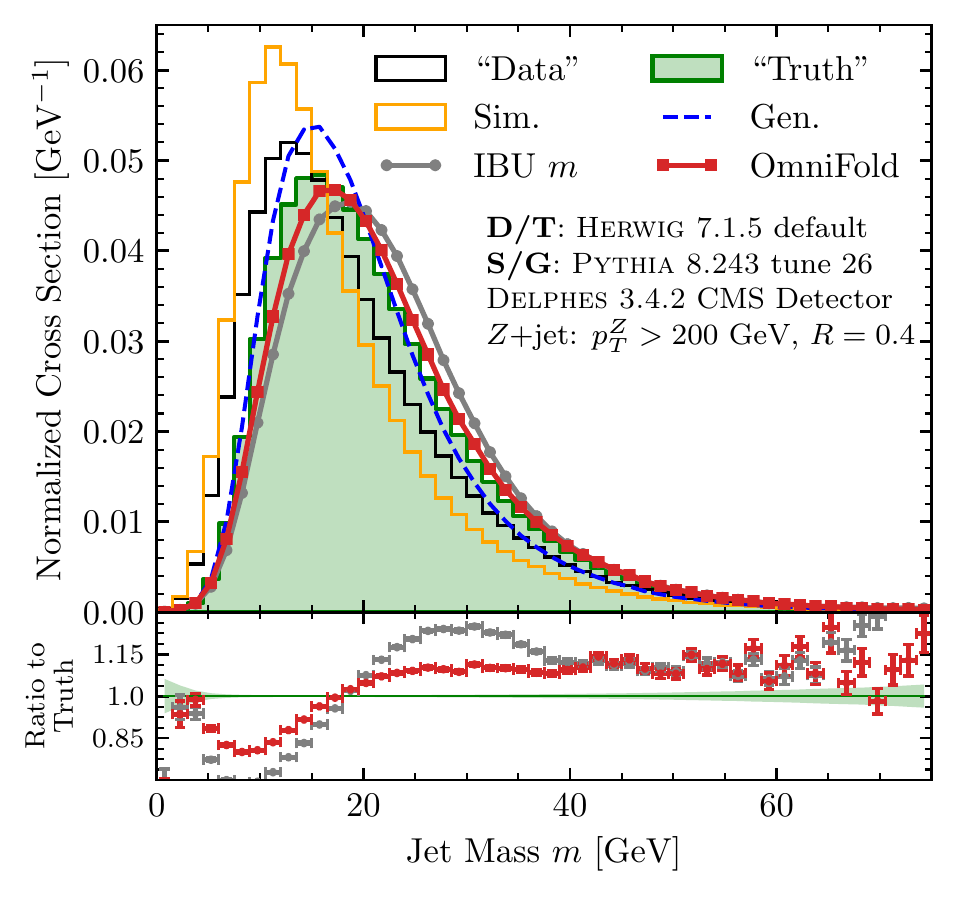}
    \includegraphics[height=0.305\textwidth]{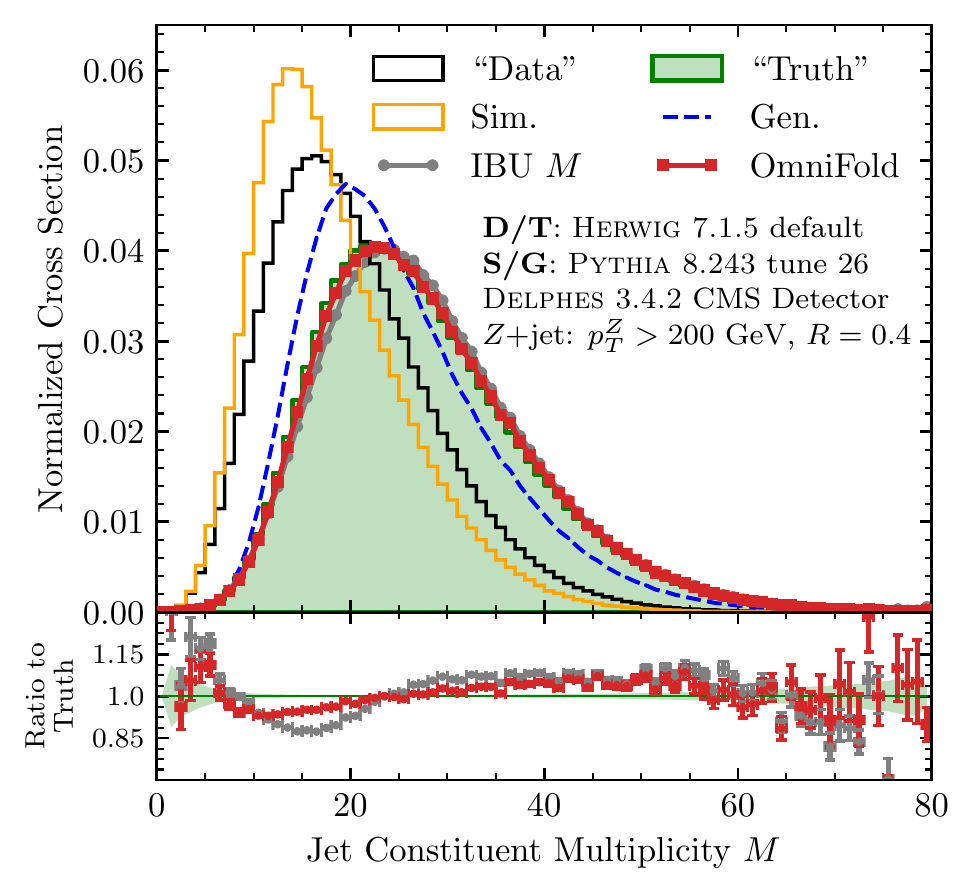}
    \includegraphics[height=0.305\textwidth]{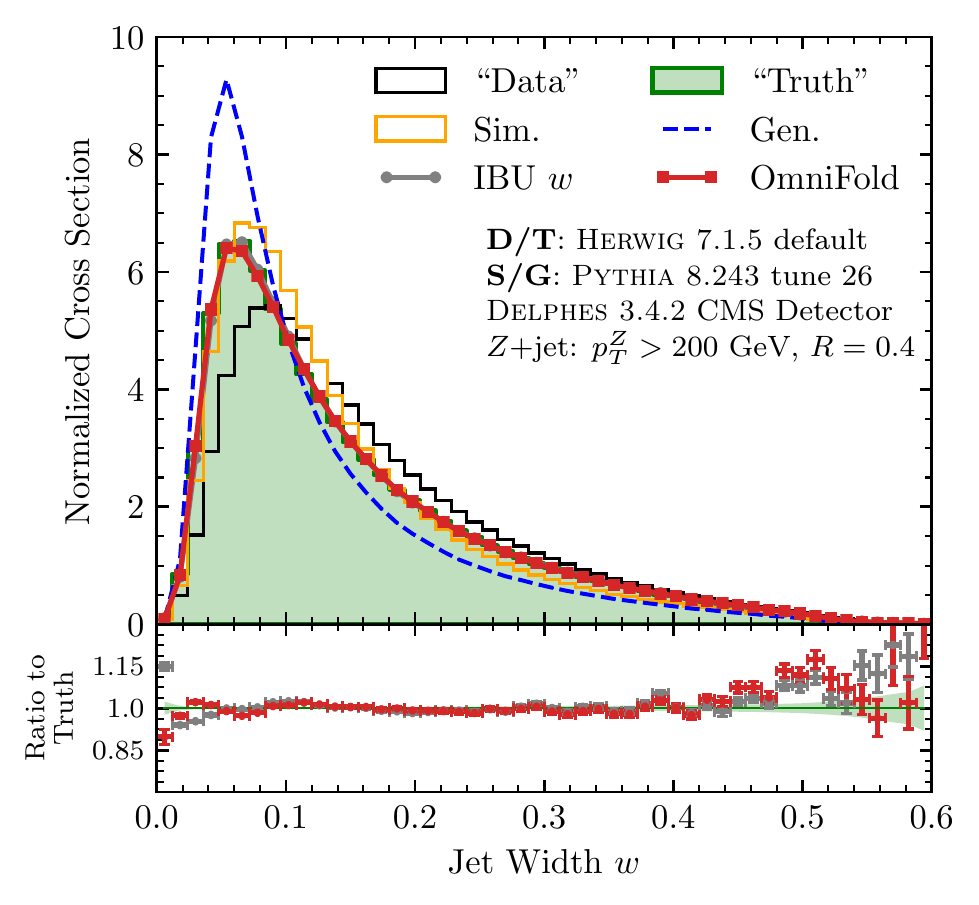}\\
    \includegraphics[height=0.305\textwidth]{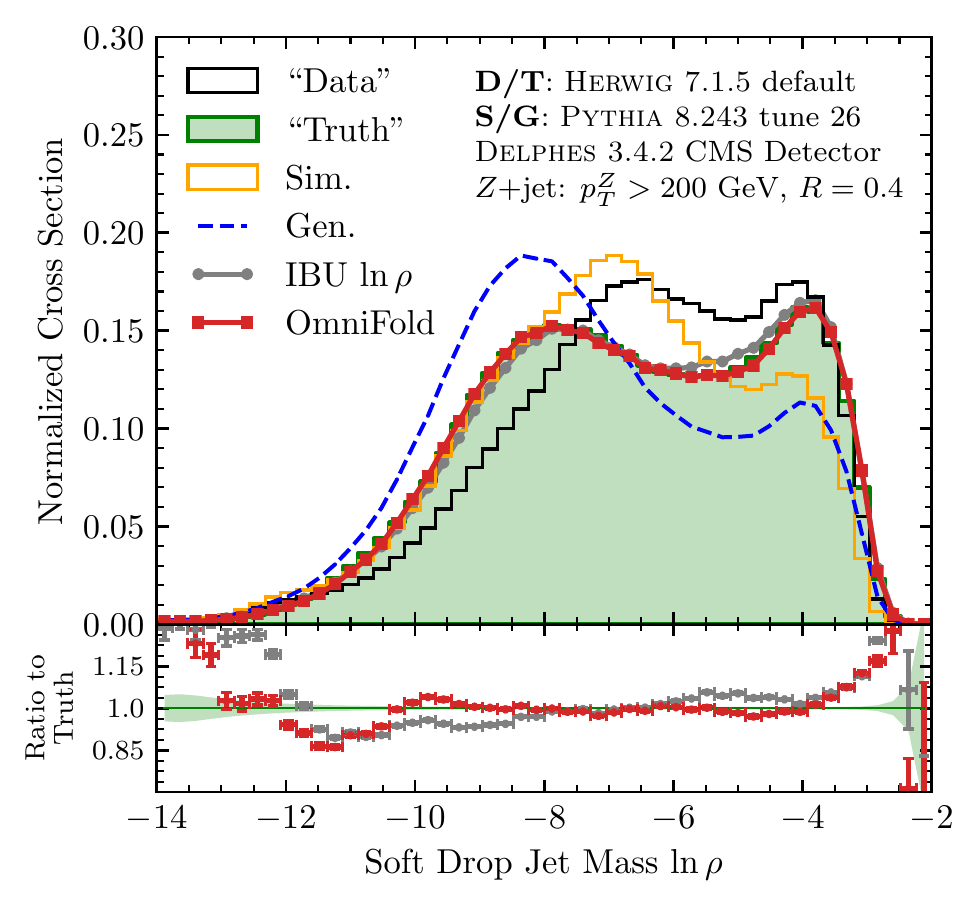}
    \includegraphics[height=0.305\textwidth]{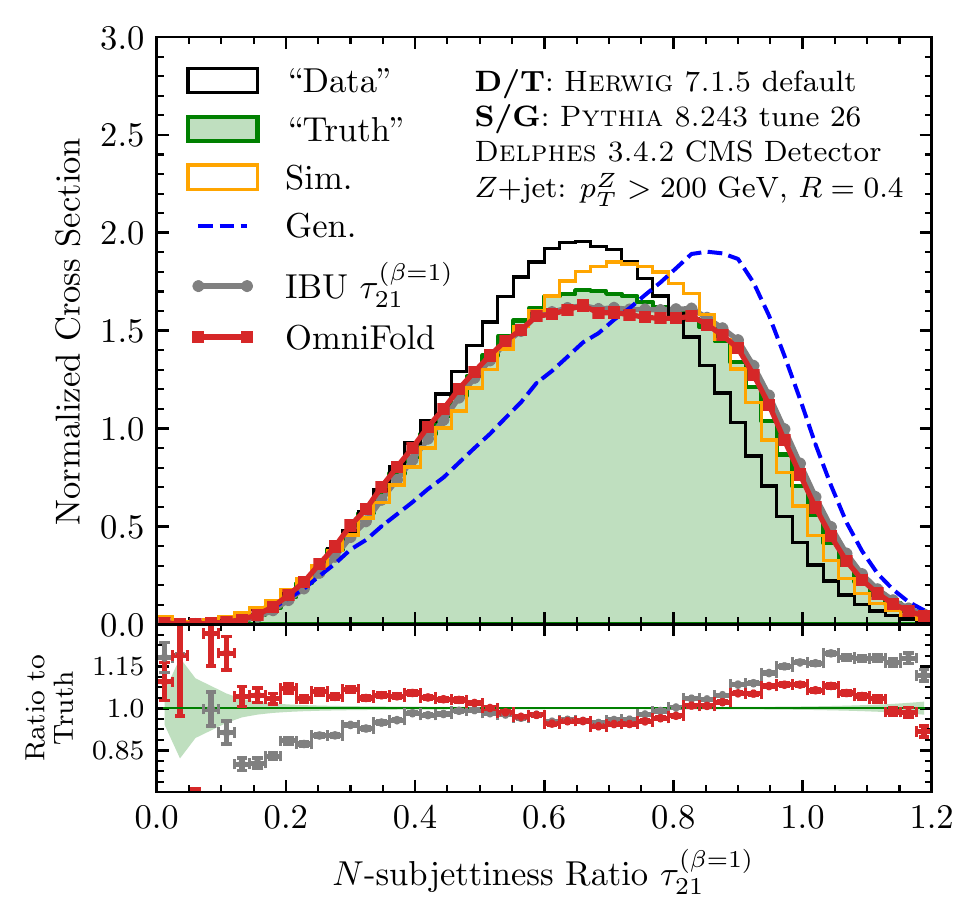}
    \includegraphics[height=0.305\textwidth]{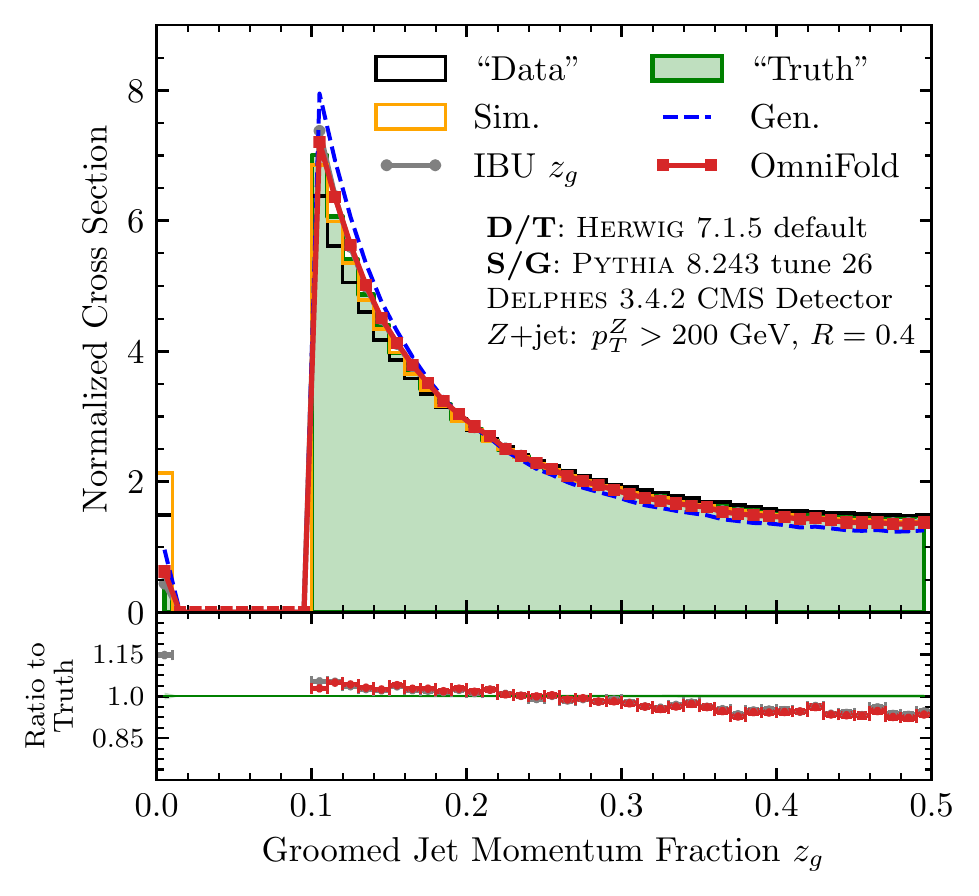}
    \caption{
    \label{fig:unfoldmultiple}
    The unfolding results for six jet substructure observables, using \Herwig~7.1.5 (``Data''/``Truth'') and \Pythia~8.243 tune 26 (Sim./Gen.), unfolded with \OmniFold and compared to IBU.
    \OmniFold matches or exceeds the unfolding performance of IBU on all of these observables.
    We emphasize that \OmniFold is a single general unfolding procedure, whereas unfolding with IBU must be done observable by observable.
    Statistical uncertainties are shown only in the ratio panel.
    }
\end{figure*}

To demonstrate the versatility and power of \OmniFold, we perform a proof-of-concept study relevant for the LHC.
Specifically, we unfold the full radiation pattern (i.e.~full phase space) of jets, which are collimated sprays of particles arising from the fragmentation and hadronization of high-energy quarks and gluons.
Jets are an ideal environment in which to benchmark unfolding techniques, since detector effects often account for a significant portion of the experimental measurement uncertainties for many jet substructure observables~\cite{Asquith:2018igt}.
With the radiation pattern unfolded, one can obtain the unfolded distribution of \emph{any} observable using \Eq{final_answer}.
Hence, this procedure can be viewed as simultaneously unfolding all observables.

Our study is based on proton-proton collisions generated at $\sqrt{s}=14$ TeV with the default tune of \Herwig~7.1.5~\cite{Bahr:2008pv,Bellm:2015jjp,Bellm:2017bvx} and Tune 26~\cite{ATL-PHYS-PUB-2014-021} of \Pythia~8.243~\cite{Sjostrand:2007gs,Sjostrand:2006za,Sjostrand:2014zea} in order to study a challenging setting where the ``natural'' and ``synthetic'' distributions are substantially different.
As a proxy for detector effects and a full detector simulation, we use the \Delphes~3.4.2~\cite{deFavereau:2013fsa} fast simulation of the CMS detector, which uses particle flow reconstruction.
Jets with radius parameter $R=0.4$ are clustered using either all particle flow objects (detector-level) or stable non-neutrino truth particles (particle-level) with the anti-$k_T$ algorithm~\cite{Cacciari:2008gp} implemented in \FastJet~3.3.2~\cite{Cacciari:2011ma,Cacciari:2005hq}.
One of the simulations (\Herwig) plays the role of ``data''/``truth'', while the other (\Pythia) is used to derive the unfolding corrections.
To reduce acceptance effects, the leading jets are studied in events with a $Z$ boson with transverse momentum $p_T^Z>200$~GeV.
After applying the selections, we obtain approximately 1.6 million events from each generator.

Any suitable machine learning architecture can be used for \OmniFold.
For this study, we use Particle Flow Networks (PFNs)~\cite{Komiske:2018cqr,DBLP:conf/nips/ZaheerKRPSS17} to process jets in their natural representation as sets of particles.
Intuitively, PFNs learn and processes a set of additive observables via $\text{PFN}(\{p_i\}_{i=1}^M) = F\left(\sum_{i=1}^M \Phi(p_i)\right)$ for an event with $M$ particles $p_i$, where $F$ and $\Phi$ are parameterized by fully-connected networks.
We specify the particles by their transverse momentum $p_T$, rapidity $y$, azimuthal angle $\phi$, and particle identification code~\cite{Tanabashi:2018oca}, restricted to the experimentally-accessible information (PFN-Ex~\cite{Komiske:2018cqr}) at detector-level.
To define separate models for Step~\ref{step1} and Step~\ref{step2}, we use the PFN architecture and training parameters of \Ref{Komiske:2018cqr} with latent space dimension $\ell = 256$, implemented in the \EnergyFlow Python package~\cite{EnergyFlow}.
Neural networks are trained with Keras~\cite{keras} and TensorFlow~\cite{tensorflow} using the Adam~\cite{adam} optimization algorithm.
The models are randomly initialized in the first iteration and subsequently warm-started using the model from the previous iteration.
20\% of the events are reserved as a validation set during training.

To investigate the unfolding performance, we consider six widely-used jet substructure observables~\cite{Larkoski:2017jix}.
The first four are jet mass $m$, constituent multiplicity $M$, the $N$-subjettiness ratio $\tau_{21}=\tau_2^{(\beta=1)}/\tau_1^{(\beta=1)}$~\cite{Thaler:2010tr,Thaler:2011gf}, and the jet width $w$ (implemented as $\tau_1^{(\beta=1)}$).
Since jet grooming~\cite{Krohn:2009th,Ellis:2009me,Ellis:2009su,Dasgupta:2013ihk,Larkoski:2014wba} is of recent interest, we also show the jet mass $\ln\rho = \ln m_\text{SD}^2/p_T^2$ and momentum fraction $z_g$ after Soft Drop grooming~\cite{Larkoski:2014wba,Dasgupta:2013ihk} with $z_\text{cut}=0.1$ and $\beta=0$.
Several of these observables are computed with the help of \textsc{FastJet Contrib 1.042}~\cite{fjcontrib}.

The unfolding performance of \OmniFold is shown in \Fig{unfoldmultiple} and compared to IBU, both with $n=5$ iterations.
We found little difference between $n=3$ and $n=5$, though \OmniFold exhibits a slight preference for more iterations.
\OmniFold succeeds in simultaneously unfolding all of these observables, achieving performance comparable to or better than IBU applied to each observable individually.
The mass is challenging for all methods as particle-type information is relevant at particle-level but is not fully known at detector-level, introducing additional prior dependence.
Though \OmniFold is unbinned, the data are only able to constrain energy and angular scales comparable to the detector resolution.

\begin{table}
  \setlength{\tabcolsep}{0pt}
  \begin{tabular}{l|>{\centering}p{3.25em}|>{\centering}p{3.25em}|>{\centering}p{3.25em}|>{\centering}p{3.25em}|>{\centering}p{3.25em}|>{\centering}p{3.25em}}\hline\hline
     & \multicolumn{6}{c}{Observable} \tabularnewline \hline\hline
    \multicolumn{1}{c|}{Method} & $m$ & $M$ & $w$ & $\ln\rho$ & $\tau_{21}$ & $z_g$ \tabularnewline \hline\hline
    \textsc{OmniFold} & \cellcolor[rgb]{0.900,1.000,0.950}\textbf{2.77} & \cellcolor[rgb]{0.482,0.666,0.491}\textbf{0.33} & \cellcolor[rgb]{0.425,0.620,0.428}0.10 & \cellcolor[rgb]{0.486,0.669,0.495}\textbf{0.35} & \cellcolor[rgb]{0.531,0.705,0.545}0.53 & \cellcolor[rgb]{0.569,0.735,0.585}0.68 \tabularnewline
    \textsc{MultiFold}\, & \cellcolor[rgb]{0.900,1.000,0.950}3.80 & \cellcolor[rgb]{0.624,0.779,0.646}0.89 & \cellcolor[rgb]{0.424,0.619,0.426}\textbf{0.09} & \cellcolor[rgb]{0.492,0.674,0.501}0.37 & \cellcolor[rgb]{0.465,0.652,0.471}\textbf{0.26} & \cellcolor[rgb]{0.437,0.630,0.441}\textbf{0.15} \tabularnewline
    \textsc{UniFold} & \cellcolor[rgb]{0.900,1.000,0.950}8.82 & \cellcolor[rgb]{0.767,0.893,0.803}1.46 & \cellcolor[rgb]{0.435,0.628,0.439}0.15 & \cellcolor[rgb]{0.547,0.718,0.562}0.59 & \cellcolor[rgb]{0.676,0.821,0.704}1.11 & \cellcolor[rgb]{0.547,0.718,0.562}0.59 \tabularnewline
    \hline
    IBU & \cellcolor[rgb]{0.900,1.000,0.950}9.31 & \cellcolor[rgb]{0.776,0.901,0.814}1.51 & \cellcolor[rgb]{0.425,0.620,0.428}0.11 & \cellcolor[rgb]{0.578,0.743,0.596}0.71 & \cellcolor[rgb]{0.676,0.821,0.704}1.10 & \cellcolor[rgb]{0.492,0.674,0.501}0.37 \tabularnewline
    \hline
    Data & 24.6 & 130 & 15.7 & 14.2 & 11.1 & 3.76 \tabularnewline
    Generation & 3.62 & 15 & 22.4 & 19 & 20.8 & 3.84 \tabularnewline
    \hline\hline
  \end{tabular}
\caption{\label{tab:ufcomp}
The unfolding performance of \OmniFold, \MultiFold, and \UniFold on six jet substructure observables, compared to IBU.
The performance is quantified by the triangular discriminator~\cite{850703,Gras:2017jty,Bright-Thonney:2018mxq} $\Delta(p,q)=\frac12 \int d\lambda \frac{(p(\lambda) - q(\lambda))^2}{p(\lambda)+q(\lambda)}$ ($\times 10^3$) between the unfolded and truth-level (binned) histograms.
Also shown are the distances from data (no unfolding) and generation (the prior).
The best unfolding method for each observable is shown in bold.
All methods perform well, with \OmniFold providing consistently good performance.
}
\end{table}

Statistical uncertainties from the prior distribution are shown in the bottom panels of \Fig{unfoldmultiple}, holding the unfolding procedure (i.e.\ response matrix and reweighting) fixed.
For this proof-of-concept study, we do not show systematic uncertainties, though the procedure for deriving them is the same as for IBU.
Non-closure and modeling uncertainties can be derived in the standard way by testing the procedure on different Monte Carlo samples and comparing the results to the known ``truth'' distributions.
(We checked that \OmniFold satisfies technical closure when \Pythia is unfolded to itself.)
Experimental systematic uncertainties can be obtained by varying the relevant effects and repeating the unfolding procedure.
Like other unfolding procedures, \OmniFold cannot improve the results in phase-space regions that are unconstrained by observed quantities.
It can, however, improve the performance if the full phase space contains auxiliary features relevant for the detector response.
To capitalize on this full phase-space approach, it is essential that the detector simulation properly describes these features and that systematic uncertainties are estimated using a high-dimensional approach~\cite{Nachman:2019yfl,Nachman:2019dol}.

\begin{figure}[t]
    \centering
    \includegraphics[width=0.925\columnwidth]{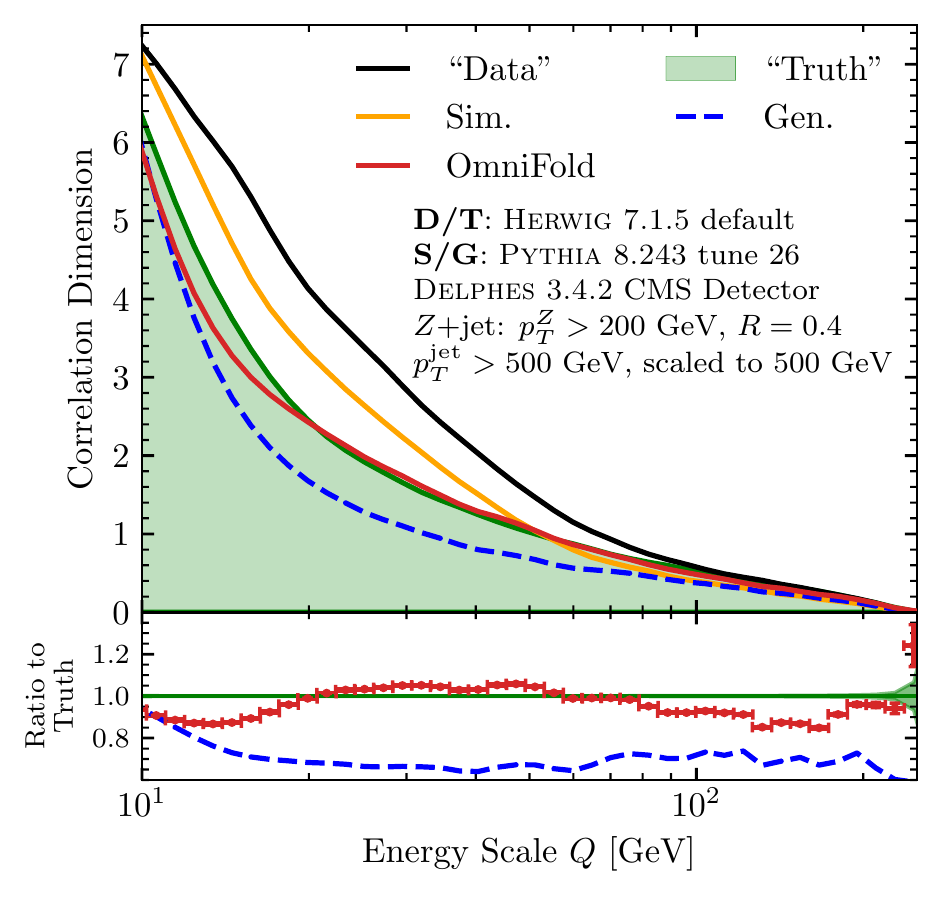}
    \caption{
    The correlation dimension of the space of jets, unfolded with \OmniFold.
    The unfolded results closely match the truth-level dimension over most of the energy range, tending toward the prior in the more difficult phase space region at low $Q$.
    Unfolding a complicated statistic such as the correlation dimension is challenging with standard methods.
    }
    \label{fig:corrdim}
\end{figure}

To highlight the flexibility of our unfolding framework, we study variations of \OmniFold, where the available information is varied by controlling the inputs:
\begin{itemize}
    \item \UniFold: A single observable as input.  This is an unbinned version of IBU.
    \item \MultiFold: Many observables as input.  Here, we use the six jet substructure observables in \Fig{unfoldmultiple} to derive the detector response.
    \item \OmniFold: The full event (or jet) as input, using the full phase space information.
\end{itemize}
The unfolding performance of each method on our six substructure observables is tabulated in \Tab{ufcomp} and compared to IBU.
The \UniFold and \MultiFold implementations both use dense networks with three layers of one hundred nodes each and a two-node output layer.
We see good unfolding performance across all methods, and even though \OmniFold is not directly trained on these six observables, it performs comparably to or better than \MultiFold.
While the detector response depends on the jet rapidity, we checked that \MultiFold did not significantly benefit from including the rapidity, though doing so could be important in a real experimental context.
In general, additional information can be included and the unfolding procedure can be repeated, with the final model chosen as the one with the best detector-level agreement with the data.

%
Since \OmniFold unfolds the full radiation pattern, it can be used to probe new, physically-interesting quantities that are challenging to unfold with existing methods.
One example is the recently-proposed correlation (fractal) dimension of jets~\cite{Komiske:2019fks,Komiske:2019jim}, which is a function of the energy scale $Q$.
This complicated statistic is defined by pairwise metric distances between jet radiation patterns, falling outside of the purview of single-observable unfolding techniques.
Within our jet samples, we restrict to energetic jets with $p_T^{\rm{jet}}>500$ GeV, boosted to the origin of the rapidity-azimuth plane, and with constituents rescaled to have $p_T$ summing to 500 GeV.
The correlation dimensions of these jets, both before and after applying \OmniFold, are shown in \Fig{corrdim}.
The unfolded results match the true distribution over a wide range of $Q$ values, with residual prior dependence seen in at low $Q$ (i.e.\ the infrared) where jets have a higher dimensionality and detector effects have a larger impact,  thus making the unfolding problem more difficult.
More broadly, \OmniFold opens the door to going beyond per-event collider observables towards more nuanced or intricate measurements of the data.

In conclusion, we have presented a potentially transformative unfolding paradigm based on iteratively reweighting a set of simulated events with machine learning.
Our \OmniFold approach allows an entire dataset to be unfolded using all of the available information, avoiding the need for binning and restricting to single observables.
We have demonstrated the power of this method in a (simulated) case of interest by unfolding the full radiation pattern of jets, paving the way for significant advances in jet substructure at the LHC.
Our unfolding framework allowed us to go beyond per-event observables towards unfolding more complex dataset statistics, such as fractal dimensions of the space of jets.
Going even further, (unsupervised) machine learning models may be trained directly at particle-level by using the unfolded and weighted dataset, which is a fascinating avenue for further exploration.
These advances have broad applicability beyond particle physics in domains where deconvolution or unfolding is used, such as image-based measurements and quantum computation~\cite{qisunfolding}.
To enable future unfolding studies and developments, we have made our code and jet datasets publicly available~\cite{omnifoldcode,omnifoldsamples}, including two additional tunes of \Pythia beyond those presented here.
Finally, our reweighting-based unfolding strategy allows for new observables to be measured long after the unfolding is carried out, which can significantly empower future public and archival collider data analyses~\cite{OpenNotEnough}.

\acknowledgments

We are grateful to Gabriel Collin, Christian Herwig,  Andrew Larkoski, Matthew LeBlanc, Salvatore Rappoccio, Jennifer Roloff, Matthew Schwartz, Daniel Whiteson, and Mike Williams for helpful comments and discussions.
We also thank the referees for their valuable suggestions, including the clarification of the statistical framing of the method.
The authors benefited from the hospitality of the Harvard Center for the Fundamental Laws of Nature.
This work was partially completed at the Aspen Center for Physics, which is supported by National Science Foundation grant PHY-1607611.
AA and BN were supported by the U.S.~Department of Energy (DOE), Office of Science under contract DE-AC02-05CH11231. 
PTK, EMM, and JT were supported by the DOE Office of Nuclear Physics under Grant No.\ DE-SC0011090 and the DOE Office of High Energy Physics under Grant Nos.\ DE-SC0012567 and DE-SC0019128.
JT is additionally supported by the Simons Foundation through a Simons Fellowship in Theoretical Physics.
Cloud computing resources were provided through a Google Cloud allotment from the MIT Quest for Intelligence.
BN would like to thank NVIDIA for providing Volta GPUs for neural network training.

\appendix

~\\

\emph{Emily Dickinson, \#975}

\begin{quote}
The Mountain sat upon the Plain \\
In his tremendous Chair --\\
His observation omnifold,\\
His inquest, everywhere --

The Seasons played around his knees\\
Like Children round a sire --\\
Grandfather of the Days is He\\
Of Dawn, the Ancestor --
\end{quote}
\hfill  \includegraphics[width=.3in]{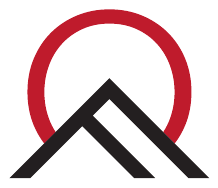}

\onecolumngrid

\section{Appendix: \OmniFold as a Maximum Likelihood Estimate}

In this Appendix, we review the statistical underpinnings of Iterative Bayesian Unfolding (IBU)~\cite{DAgostini:1994fjx} as well as \OmniFold and confirm that they converge to the maximum likelihood estimate of the true particle-level distribution.
This discussion serves to clarify the statistical formulation of unfolding, as well as to provide a derivation of the correctness of \OmniFold.
We follow the overall spirit of \Ref{shepp1982maximum}, keeping the formulation general and unbinned, working in the asymptotic limit of large amounts of data.
We seek to find the truth-level reweighting $\nu(t)$ of the synthetic particle-level distribution $p_\text{Gen.}(t)$ that maximizes the likelihood of observing the measured data.

The (log) likelihood of a given reweighting to produce the observed data is:
\begin{align}\label{eq:LL}
\text{LL}[\nu] & = \int dm\, p_\text{Data}(m)\,\ln\left(\int dt'\, p(m|t') \, \nu(t') \, p_\text{Gen.}(t') \right) - \lambda \left(\int dt'  \nu(t') \, p_\text{Gen.}(t') - 1\right).
\end{align}
Here, $p_\text{Data}(m)$ is the measured distribution of data at detector level.
The quantity $p(m|t)$ captures the detector response, i.e.\ the distribution of the detector-level information $m$ produced by the particle-level information $t$.
We take the detector response to be accurately modeled in the synthetic dataset, $p(m|t) = p_\text{Data$|$Truth}(m|t) = p_\text{Sim.$|$Gen.}(m|t)$, one of the standard assumptions of unfolding.
The last term is a Lagrange multiplier constraint to ensure that the particle-level reweighting $\nu(t)$ of the synthetic distribution $p_\text{Gen.}(t)$ yields a normalized distribution.

To maximize the likelihood, we vary it with respect to the reweighting function $\nu$:
\begin{equation}\label{eq:varyLL}
\frac{\delta \text{LL}}{\delta \nu(t)} = \int dm \, p_\text{Data}(m) \, \frac{p(m|t) \, p_\text{Gen.}(t)}{\int dt'\,p(m|t') \, \nu(t')\, p_\text{Gen.}(t')} - \lambda \, p_\text{Gen.}(t) = 0.
\end{equation}
Integrating this equation equation against $\int dt\, \nu(t)$ and applying the normalization condition yields that $\lambda=1$.
The stationary condition in \Eq{varyLL} results in a maximum of the likelihood because the second variation is non-positive and therefore the functional is concave:
\begin{equation}\label{eq:varyLL2}
\frac{\delta^2 \text{LL}}{\delta \nu(t_0) \, \delta \nu(t_1)} = -\int dm \, p_\text{Data}(m) \, \frac{p(m|t_0) \, p_\text{Gen.}(t_0) \, p(m|t_1) \, p_\text{Gen.}(t_1)}{\left(\int dt'\,p(m|t') \,\nu(t') \, p_\text{Gen.}(t')\right)^2} \le 0.
\end{equation}

To connect the maximum likelihood strategy to the \OmniFold and IBU methods, we can multiply the stationary condition in \Eq{varyLL} on both sides by $\nu(t)$ to obtain an equation satisfied by the optimal reweighting function $\nu^*(t)$:
\begin{equation}\label{eq:LLcond}
\nu^*(t) \, p_\text{Gen.}(t) = \int dm \, p_\text{Data}(m) \, \frac{p(m|t) \, \nu^*(t) \, p_\text{Gen.}(t)}{\int dt'\, p(m|t') \, \nu^*(t') \, p_\text{Gen.}(t')}.
\end{equation}
If we were to replace $\nu^*(t)$ on the left-hand side of \Eq{LLcond} by $\nu_n(t)$ and on the right-hand side by $\nu_{n-1}(t)$, we would obtain the update rule for \OmniFold, with the discrete version corresponding to IBU:
\begin{equation}\label{eq:iter}
\nu_n(t) \, p_\text{Gen.}(t) = \int dm \, p_\text{Data}(m) \, \frac{p(m|t) \, \nu_{n-1}(t) \, p_\text{Gen.}(t)}{\int dt'\, p(m|t') \, \nu_{n-1}(t') \, p_\text{Gen.}(t')}.
\end{equation}

To see that \Eq{iter} indeed causes the likelihood to increase, we will show that it is a consequence of a generalized expectation-maximization (EM) algorithm~\cite{10.2307/2984875,wu1983,doi:10.1080/01621459.1985.10477119,Kuusela2012StatisticalII} in which the likelihood increases at each step.
We use the particle-level information $t$ as the unobserved latent variables.
By Theorem 1 of \Ref{10.2307/2984875}, we only need to show that the expected complete-data (log) likelihood $Q$ increases from one choice of reweighting $\nu_{n-1}(t)$ to another $\nu_n(t)$:
\begin{equation}\label{eq:completeLL}
Q(\nu_n|\nu_{n-1})=\int dm\,p_\text{Data}(m) \int dt\,p(t|m,\nu_{n-1})\ln p(t,m|\nu_n).
\end{equation}
Manipulating the argument of \Eq{completeLL} using conditional probabilities and Bayes' rule, we have:
\begin{equation}\label{eq:Q}
Q(\nu_n|\nu_{n-1})=\int dm\,p_\text{Data}(m) \int dt\,\frac{p(m|t) \, \nu_{n-1}(t) \, p_\text{Gen.}(t)}{\int dt'\,p(m|t') \, \nu_{n-1}(t') \, p_\text{Gen.}(t')}\ln\big(\nu_n(t) \, p_\text{Gen.}(t)\big) + \text{const.},
\end{equation}
where the response $p(m|t)$ is independent of $\nu$ and $p(t|\nu) = \nu(t) \, p_\text{Gen.}(t)$.
The constant term is independent of $\nu_n$ and so will not contribute to our maximization over $\nu_n$.

We can maximize \Eq{Q} with respect to $\nu_n(t)$, including the Lagrange multiplier constraint to enforce normalization, by taking the derivative and setting it equal to zero.
This collapses the integral over $t$ and leads exactly to \Eq{iter}.
Thus the choice of $\nu_n(t)$ via \Eq{iter} increases $Q$, and so the log likelihood also increases by the properties of the generalized EM algorithm.
While the maximum may not be unique due to null directions in \Eq{varyLL2}, a global maximum of the likelihood is attained due to its concavity, with the precise $\nu_\infty(t)$ dictated by the choice of initial distribution $p_\text{Gen.}(t)$.
If the response ``matrix'' $p(m|t)$ is invertible, this procedure converges to the true solution.
Further, terminating the algorithm after a finite number of iterations introduces regularization by reducing the variance of the estimator at the cost of increased bias via prior dependence.

Thus, \OmniFold provides an unbinned, machine-learning-based strategy to estimate the maximum likelihood true particle-level distribution given the observed detector-level data.
Of course, in the presence of statistical fluctuations, the maximum of the likelihood may not be the desired solution, so in practice, one regularizes the unfolding procedure by performing a finite number of iterations.

\clearpage
\twocolumngrid
\bibliography{myrefs}

\end{document}